# Spin-orbit coupling within tightly focused circularly polarized spatiotemporal vortex wavepacket

Jian Chen[1], Lihua Yu[1], Chenhao Wan[1,2], and Qiwen Zhan[1,*]

[1]*School of Optical-Electrical and Computer Engineering, University of Shanghai for Science and Technology, Shanghai 200093, China*

[2]*School of Optical and Electronic Information, Huazhong University of Science and Technology, Wuhan, Hubei 430074, China*
*Corresponding Author: qwzhan@usst.edu.cn

**Abstract**

Spin-orbital coupling and interaction as intrinsic light fields characteristics have been extensively studied. Previous studies involve the spin angular momentum (SAM) carried by circular polarization and orbital angular momentum (OAM) associated with a spiral phase wavefront within the beam cross section, where both the SAM and OAM are in parallel with the propagation direction. In this work, we study a new type of spin-orbital coupling between the longitudinal SAM and the transverse OAM carried by a spatiotemporal optical vortex (STOV) wavepacket under tight focusing condition. Intricate spatiotemporal phase singularity structures are formed when a circularly polarized STOV wavepacket is tightly focused by a high numerical aperture objective lens. For the transversely polarized components, phase singularity orientation can be significantly tilted away from the transverse direction towards the optical axis due to the coupling between longitudinal SAM and transverse OAM. The connection between the amount of rotation and the temporal width of the wavepacket is revealed. More interestingly, spatiotemporal phase singularity structure with a continuous evolution from longitudinal to transverse orientation through the wavepacket is observed for the longitudinally polarized component. These exotic spin-orbit coupling phenomena are expected to render new effects and functions when they are exploited in light matter interactions.

**Introduction**

It's well known that a photon can carry linear momentum along the propagation direction, spin angular momentum (SAM) with regard to polarization state, and orbital angular momentum (OAM) that induces the formation of optical vortices[1]. In the widely studied optical vortices in spatial domain, also colloquially called vortex beams, the longitudinal OAMs reside parallel with the axis of beams, and the null and spiral phase pattern can only be observed in the cross-section of the beams[2]. With the extensive investigations of vortex beams and longitudinal OAMs, there have been enormous interests in the spin-orbit interactions (SOI) of light. Generally speaking, the optical SAM and OAM are different rotation degrees of freedom that remain independent in homogenous, isotropic media within the limit of paraxial approximation[3], however the two angular momenta can be coupled in particular circumstances such as tightly focusing of circularly polarized beams in homogenous isotropic media[4], light-matter interaction in inhomogeneous[5], anisotropic[6] or structured[7] media. SOI has brought several intriguing optical phenomena including spin-Hall effect of light[8-10], spin-controlled shaping of light[11], interconversion between OAM and SAM of photons under strongly focusing[12,13], etc.

While many works about characteristics of SOI related to longitudinal OAM have been carried out and applied in precise metrology[14,15], optical tweezers[16], optomechanical systems[17] and so on, there have been no report on the phenomena of SOI associated with transverse OAM due to the complex natures from spatiotemporal optical vortices (STOVs) whose null in intensity and helical phase are conceived in spatiotemporal domain[18,19]. In recent years, several seminal researches on STOVs and transverse OAMs have been published such as the propagation in free space[20], controllable generation methods[21], the process of strongly focusing[22], which broadly unveil natures and provide accessible means for the generation, manipulation, and observation of STOVs. For such a new transverse OAM state of light, it is reasonable to expect that novel phenomena might arise between the SAM and OAM.

In this work, we study the circularly polarized STOV carrying transverse OAM under tightly focusing condition to offer insights into the interaction between the SAM and the transverse OAM. The transversely polarized components present a ring-shaped profile, whose phase singularity orientation can be tilt towards the optic axis away from the transverse direction owing to the coupling between the longitudinal SAM and transverse OAM. Additionally, the amount of rotation is connected to the temporal half-width of the wavepacket. For the longitudinally polarized component, spatiotemporal phase singularity experiences an evolution from longitudinal to transverse orientation through the wavepacket.

**Results and discussion**

**Pre-conditioning of incident wavepacket**

Without losing generality, we focus our discussion on the right-handed circularly polarized STOV with topological charge of +1, which can be expressed as:

$$\mathbf{E}_{+1}(x,y,t) = \frac{\sqrt{2}}{2}\begin{pmatrix}\vec{e}_x \\ i\vec{e}_y\end{pmatrix}(x+it)\exp\left(-\frac{x^2+y^2+t^2}{w^2}\right), \tag{1}$$

where $\vec{e}_x$ and $\vec{e}_y$ are the corresponding unit vectors along x- and y-axes, $w$ is the waist radius of Gaussian profile. In order to simplify the following discussions, the sizes of Gaussian profile in spatial domain and spatiotemporal domain are both normalized to the same $w$. Similarly, a right-handed circularly polarized incident STOV with topological charge of -1 is given by

$$\mathbf{E}_{-1}(x,y,t) = \frac{\sqrt{2}}{2}\begin{pmatrix}\vec{e}_x \\ i\vec{e}_y\end{pmatrix}(x-it)\exp\left(-\frac{x^2+y^2+t^2}{w^2}\right). \tag{2}$$

It has been demonstrated that the spatiotemporal spiral phase structure within STOV will collapse during tightly focusing through a high numerical aperture (NA) objective lens owing to a spatiotemporal astigmatism effect[22]. To prevent the spatiotemporal phase structure from collapsing, the preconditioning method similar to that used in cylindrical mode converter for the conversion from Hermite-Gaussian (HG) modes into Laguerre-Gaussian (LG) modes can be employed to produce tightly focused STOVs[23]. Since HG modes can be formed from linear superposition of LG modes, the superposition of the above two circularly polarized incident STOVs can be written as:

$$\mathbf{E}_t^{+1}(x,y,t) = \mathbf{E}_{+1}(x,y,t) + i\mathbf{E}_{-1}(x,y,t) = \frac{\sqrt{2}}{2}\begin{pmatrix}\vec{e}_x \\ i\vec{e}_y\end{pmatrix}(1+i)(x+t)\exp\left(-\frac{x^2+y^2+t^2}{w^2}\right). \tag{3}$$

With the preconditioning, the transverse OAM with topological charge of +1 in the STOV will be maintained after strongly focusing.

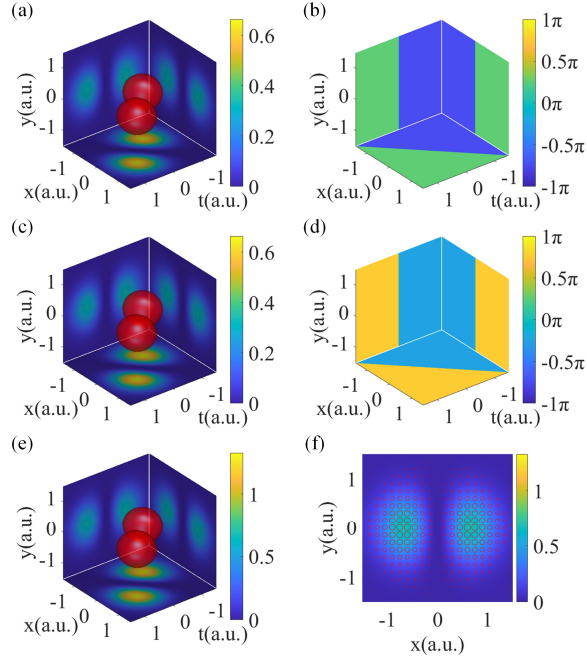

Fig. 1. Intensity and phase distributions of pre-conditioned incident STOV in the planes of $x = 0$, $y = 0$, and $t = 0$; the isosurfaces visually present the profiles at 50% of maximum intensity. (a) Intensity distributions and isosurface of the x-polarized component. (b) Corresponding phases of the x-polarized component. (c) Intensity distributions and isosurface of the y-polarized component. (d) Corresponding phases of the y-polarized component. (e) Intensity distributions and isosurface of total incident field. (f) Polarization distribution of incident wavepacket in x-y plane at $t = 0$. White represents positive phase retardation. Note that t-axis is normalized to the pulse half-width.

The preconditioned wavepacket and its each constituting component are shown in Fig. 1. Both the x- and y-polarized components are split in the spatiotemporal domain, with spatiotemporal distribution similar to that of $HG_{01}$ mode in spatial domain and a rotation of 45° with respect to the t-axis. This preconditioning of the wavepacket enables the restoration of transverse OAMs in the tightly focused wavepackets. Figs. 1(a)-1(d) also shows that the respective profiles and phase patterns of the x-polarized and the y-polarized components are identical with each other, except a phase retardation of $\pi/2$ between them. For the x-polarized component, the green area in Fig. 1(b) represents $\pi/4$ as the dark blue area of $-3\pi/4$, while the yellow and bluish in Fig. 1(d) stand for $3\pi/4$ and $-\pi/4$ respectively. The entire incident field is shown in Fig. 1(e), and its polarization distribution in x-y plane (as demonstrated in Fig. 1(f)) clearly indicates the incident wavepacket is circularly polarized in spatial domain.

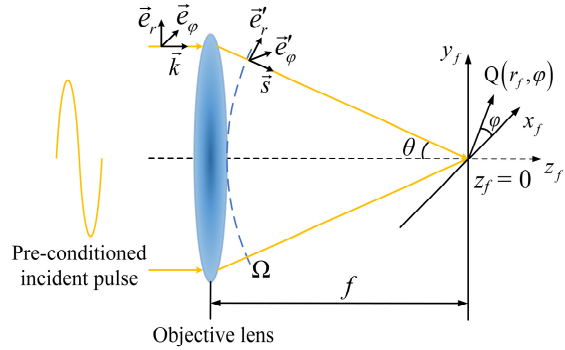

Fig. 2. Schematic diagram of the tightly focusing for STOV.

**Tight focusing of the STOV**

To calculate the focal field, a simplified model is adopted where the spatiotemporal coupling is ignored such that each temporal slice of the incident wavepacket is assumed to be focused onto its conjugate temporal location within the focal space. Chromatic and other aberrations are also ignored at the moment. Subsequently, the diffracted field in the vicinity of focal plane is calculated with the Debye integral as[24-27]

$$\mathbf{E}_f(r_f,\Phi,z_f,t) = \int_0^\alpha \int_0^{2\pi} \mathbf{P}(\theta,\varphi) B(\theta) E_\Omega(\theta,\varphi,t) \exp\{-ik[r_f \sin\theta \cos(\varphi-\Phi) + z_f \cos\theta]\} \sin\theta \, d\theta \, d\varphi, \quad (4)$$

where α is convergence angle determined by the NA of the lens, $r_f = \sqrt{x_f^2 + y_f^2}$, $z_f = 0$, $\Phi = \tan^{-1}(y_f/x_f)$, $\mathbf{P}(\theta,\varphi)$ is the polarization distribution of the refracted field, $B(\theta)$ is the apodization function of the objective lens. For aplanatic lens that obeys sine condition, $B(\theta) = \sqrt{\cos\theta}$. $E_\Omega(\theta,\varphi,t)$ is the complex amplitude distribution of the refracted field on the spherical surface Ω in Fig. 2, which can be expressed as

$$E_\Omega(\theta,\varphi,t) = (1+i)(\sin\theta \cos\varphi + t) \exp\left[-(\sin^2\theta + t^2)/w^2\right], \quad (5)$$

where the spatial size and temporal size have been normalized to the NA of the lens and the pulse half-width, respectively. In this work, NA = 0.95 and $w$ = 0.95 are used in all the following numerical simulations.

The polarization distribution $\mathbf{P}(\theta,\varphi)$ of the refracted field will influence the structure of the wavepacket in the focal region. In the case of right-handed circular polarization, $\mathbf{P}(\theta,\varphi)$ is written as:

$$\mathbf{P}(\theta,\varphi) = \begin{pmatrix} \left[(\cos\theta \cos^2\varphi + \sin^2\varphi) + i(\cos\theta \sin\varphi\cos\varphi - \sin\varphi\cos\varphi)\right]\vec{e}_x \\ \left[(\cos\theta \cos\varphi \sin\varphi - \sin\varphi\cos\varphi) + i(\cos\theta \sin^2\varphi + \cos^2\varphi)\right]\vec{e}_y \\ \left[-\sin\theta \exp(i\varphi)\right]\vec{e}_z \end{pmatrix} = \begin{pmatrix} (\cos\theta \cos\varphi - i\sin\varphi)\vec{e}_x \\ (\cos\theta \sin\varphi + i\cos\varphi)\vec{e}_y \\ (-\sin\theta)\vec{e}_z \end{pmatrix} \exp(i\varphi). \quad (6)$$

Here, the following transformation relationships are used:

$$\vec{e}_r = \cos\varphi \vec{e}_x + \sin\varphi \vec{e}_y, \quad (7)$$

$$\vec{e}_r' = \cos\theta(\cos\varphi \vec{e}_x + \sin\varphi \vec{e}_y) + \sin\theta \vec{e}_z, \quad (8)$$

$$\vec{e}_\varphi' = \vec{e}_\varphi = -\sin\varphi \vec{e}_x + \cos\varphi \vec{e}_y, \quad (9)$$

where $\vec{e}_r$ and $\vec{e}_\varphi$ are the radial and azimuthal unit vectors in the pupil plane of the lens, respectively; $\vec{e}_r'$ and $\vec{e}_\varphi'$ are the corresponding radial and azimuthal unit vectors in the image space; $\vec{e}_z$ is the longitudinal unit vector.

It is very interesting to note that a spatial vortex with topological charge of +1 appears for all three polarized components in the integral, as shown in Eq. (6). This term arises from the circular polarization (i.e. SAM) of the input polarization casted in the cylindrical coordinates. It plays an important role to understand the spin-orbital coupling in the tight focusing process.

**Transversely polarized components**

First, let's examine the transversely polarized components (i.e. the x- and y-components) for tightly focused right-handed circularly polarized STOV. In Fig. 3, the profiles of transversely polarized components seem to roughly remain what are like in scalar field in Ref. [22]. As seen in Figs. 3(a) and 3(c), both the isosurfaces of the x- and y-components visualize an elongated ring shape with a hollow core along the y-axis. According to the corresponding phase distributions in Figs. 3(b) and 3(d), clockwise helical phases in spatiotemporal domain are both observed in the x-t plane. The retardation of π/2 is also preserved between the two transverse components. This indicates that the state of polarization for the focused wavepacket is still mainly circularly polarized, as one would expect.

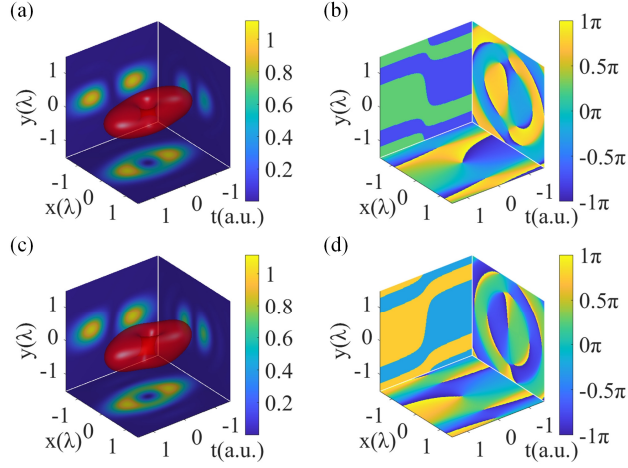

Fig. 3. Intensity and phase distributions of the x- and y-polarized components of the focused wavepacket in $x = 0$, $y = 0$, and $t = 0$ planes, the isosurfaces is depicted at 20% of its maximum intensity. (a) and (b) Intensity, phase distributions, and isosurface of the x-component. (c) and (d) Intensity, phase distributions, and isosurface of the y-component.

However, careful inspection of the phase pattern reveals more intricate phase singularity behavior due to the coupling between longitudinal SAM and intrinsic transverse OAM. From Figs. 3(b) and 3(d), it can be seen that both the x- and y-components also have a helical phase projection on the x-y plane, in addition to the helical phase projection in the x-t plane. This indicates that the phase singularity wavefront for both transverse components has a tilt angle with respect to the z-axis (t-axis). To better illustrate this, slices of the spatiotemporal intensity and phase distributions for both x- and y-polarized components at $x = 0$ plane are shown in Fig. 4. As illustrated in Figs. 4(a) and 4(c), the intrinsic OAM in the x-polarized component together with its STOV is clockwise rotated with regard to the y-axis while the counterpart in the y-polarized component is anti-clockwise rotated. It may seem that the tilt angle is relatively small. However, it should be noted that the t-axis is normalized to the pulse half-width. The tilt angle turns out to be very substantial when realistic pulse width is considered.

According to Figs. 4(b) and 4(d), the diameter of the wavepacket $2w_f = 0.64\lambda$. Assuming the pulse half-width is $\tau = 1\text{ps}$ and the wavelength is 1μm, the temporal width of the tilt in Figs. 4(b) and 4(d) is $\tau_t = 0.07\text{ps}$. The corresponding tilt angle is actually $\tan^{-1}\left[\tau_t c/(2w_f)\right] \cong 88.25°$, which means the OAM associated with this phase singularity is almost along the propagation direction. This can be understood with the following simplified picture. The cross-sectional area of the wavepacket in the spatiotemporal plane where the transverse OAM resides can be approximately calculated as $\pi\tau c w_f$ while the counterpart in spatial plane carrying SAM is $\pi w_f^2$. The ratio between the cross-sectional area of OAM and SAM is $\tau c w_f$, rendering significantly lower OAM density compared to SAM density. Thus, the total angular momentum density within the focused wavepacket is significantly skewed by the longitudinal

SAM. If an x- or y-polarizer is employed, the transmitted field will have an OAM orientation dragged nearly in parallel with the propagation axis due to the coupling and conversion of longitudinal SAM with the initially transverse OAM.

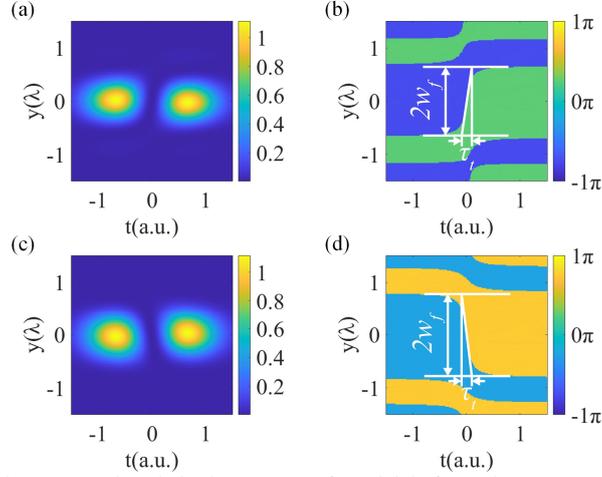

Fig. 4. Intensity and phase slices of the transversely polarized components from tightly focused STOV at $x = 0$. (a) Intensity slice of the x-polarized component. (b) Corresponding phase of the x-polarized component. (c) Intensity slice of the y-polarized component. (d) Corresponding phase of the y-polarized component. The length of the tilt boundary is dimensioned.

The analysis above also points to an effective way to tune the orientation of the OAM carried by the x- or y-polarized components through changing the pulse width, which in turn adjust the relative density between the longitudinal SAM and the transverse OAM. For example, if the pulse half-width is reduced to 10fs, the transverse OAM density will dominate over the SAM density and the tilt angle will be reduced to 18.15°. This angle can be further reduced by using even shorter pulse. However, it should be noted that the focusing model is extremely simplified and much more factors need to be considered as the pulse width approaches a few cycles of the optical oscillation. Nevertheless, the above analysis clearly showed a very unique spin-orbital coupling between the longitudinal SAM and transverse OAM under tight focusing condition. This spatiotemporal spin-orbital coupling offers capability of creating OAM with nearly arbitrary spatial orientation.

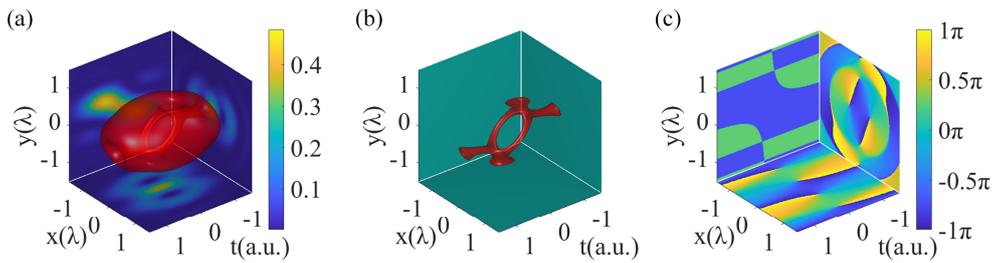

Fig. 5. (a) Intensity distributions of the z-component in the focused STOV in $t = 0, x = 0, y = 0$ planes, the isosurface is depicted at 5% of its maximum intensity. (b) Extracted hollow structure to illustrate the evolution of the phase singularities within the wavepacket. (c) Corresponding phase distributions. See Supplementary Movie 1 for animation of phase singularities evolution in different perspective.

**Longitudinally polarized component**

The z-polarized component of the focused wavepacket exhibits much more intricate phase singularity structure in the spatiotemporal domain. From Eqs. (4) and (6), the integral for the z-polarized component includes a longitudinal phase singularity converted from SAM and a transverse phase singularity embedded in the STOV. This is very different from the circularly polarized LG beam in tightly focusing of which the topological charges of the respective longitudinal

OAMs can be superimposed linearly[17]. The interaction between the longitudinal contribution from spin and the transverse contribution from the STOV produces a continuous evolution of the phase singularity structure as shown in Fig. 5(a). At the leading edge of the incident wavepacket, the phase in the x-y plane is relatively flat and the contribution to phase singularity is mainly from the longitudinal one as shown in Eq. (6), leading to a longitudinally oriented phase singularity. As the integral steps through different temporal slices, the phase pattern attributed to the STOV starts to interact with the longitudinal OAM and lead to a gradual evolution of the phase singularity orientation. Near the center of the wavepacket, the phase singularity becomes transverse and extend to the surface of the wavepacket. Subsequently the phase singularity splits into two, circumvent the core of the wavepacket and join again to form another transversely oriented phase singularity on the other side of the wavepacket. Then the orientation of the phase singularity gradually turns into longitudinal as the time slice approaching the end of the wavepacket. The complete evolution for the core of the phase singularity is shown in Fig. 5(b). Phase patterns in $t = 0$, $x = 0$, $y = 0$ planes are also shown in Fig. 5(c) to further reveal the evolution of the phase singularity through the spatiotemporal volume space. Supplementary Movies 2, 3, 4, exhibiting animation of slices along the t-, x-, and y-axes respectively, are provided for detailed evolution of the vortices and phase singularities on the three orthogonal planes.

**Polarization distributions of focused wavepacket**

The intensity distributions of the total focused wavepacket with superimposed polarization projections in corresponding planes are shown in Fig. 6. As shown in Fig. 6(a), the shape of the wavepacket is similar to those of the transversely polarized components because their intensities are almost two times higher than that of the z-polarized component. For the same reason, the polarization with regard to the x- and y-polarized components in the x-y plane roughly remains circularly polarized, as shown in Fig. 6(b). However, as shown in Figs. 6(c) and 6(d), the polarization distributions in x-t and y-t planes are mainly elliptical, since the intensity of the z-component is much smaller than those of the two transverse components. On the other hand, there almost only exist the z-polarized component at the center area of the focused wavepacket, where the intensities of the transversely polarized components are nearly zero. Thus, the center area is linearly polarized along the z-direction, as shown in Figs. 6(c) and 6(d).

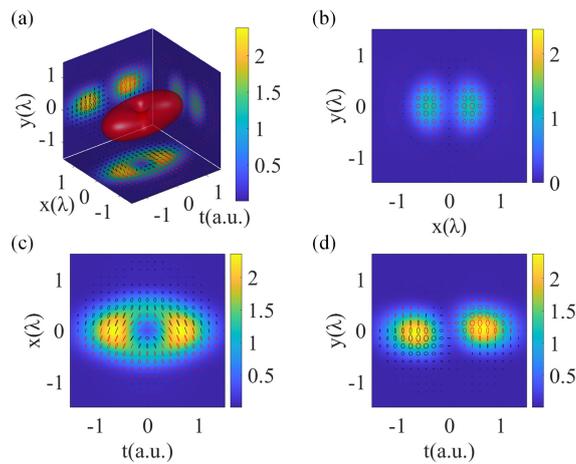

Fig. 6. (a) Intensity and polarization distributions of tightly focused circularly polarized STOV in $t = 0$, $x = 0$, $y = 0$ planes; the isosurface is depicted at 20% of maximum intensity. Red represents positive phase retardation and black indicates the negative. (b) Intensity and polarization distribution in x-y plane at $t = 0$. (c) Intensity and polarization distribution in x-t plane at $y = 0$. (d) Intensity and polarization distribution in y-t plane at $x = 0$. Note t-axis is normalized to the pulse half-width.

## Conclusions

In conclusion, we study the tight focusing of circularly polarized STOV wavepacket, offering insights into the spin-orbit interaction phenomena associated with transverse OAM for the first time. For the transversely polarized components, substantial tilt of the phase singularity orientation is found due to the strong coupling between transverse OAM and longitudinal SAM. The tilt angle can be conveniently controlled by the temporal width of the wavepacket. Much more complicated phase singularity structures are discovered for the longitudinally polarized component of the focal field. Continuous evolution of phase singularity from purely longitudinal to purely transverse orientation can occur inside the z-component of the focused wavepacket with a closed knot kind of connections in the center of the wavepacket. When interacts with structure materials, these unique features of the spin-orbital coupling may render distinctively different response, especially if the sample selectively responds to certain specific state of polarization. For example, quantum emitters coupled with plasmonic structures are found to be strongly coupled with the z-components that have been exploited in directional emission of photons with specific SAM and/or OAM states[28-30]. Using the focused circularly polarized STOV as the excitation, it is then possible to create photon emissions with ultrafast dynamically modulated SAM and/or OAM states. It should be noted that a simple vectorial focusing model is used in this study and many practical aspects, such as chromatic aberration, lens aberration and spatiotemporal coupling arising from ultrashort pulses, are not considered. It is expected that much richer spin-orbital interactions will be discovered when these contributions are further incorporated in the future studies. The unique spin-orbital interaction offered by focused circularly polarized STOVS may find potential applications in optical manipulation, photon emission with tailored properties, light-matter interactions and plasma physics and so on.

## Acknowledgements

This research is supported by National Natural Science Foundation of China (92050202, 61805142, 61875245); Shanghai Science and Technology Committee (19060502500); Shanghai Natural Science Foundation (20ZR1437600).

## Competing interests

The authors declare no competing interests.

## Author Contributions

J. C. and L. Y. conducted the numerical simulations, Q. Z. supervised the project. All authors analyzed the results and contributed to the writing.

## Data availability

Simulated data and source code are available from corresponding author upon reasonable request.

## Code availability

Source code written in Matlab for numerical simulation is available from the corresponding author upon reasonable request.